\documentclass[twocolumn,showpacs,prx]{revtex4-1}%
\usepackage{amsfonts}
\usepackage{amsmath}
\usepackage{amssymb}
\usepackage{subfigure}
\usepackage{graphicx}
\usepackage{txfonts}%
\providecommand{\U}[1]{\protect\rule{.1in}{.1in}}

\begin{document}
\title{Tilted Klein tunneling across atomically sharp interfaces}
\author{Shu-Hui Zhang$^{1}$}
\author{Wen Yang$^{2}$}
\email{wenyang@csrc.ac.cn}
\author{Kai Chang$^{3}$}
\affiliation{$^{1}$College of Science, Beijing University of Chemical Technology, Beijing,
100029, China}
\affiliation{$^{2}$Beijing Computational Science Research Center, Beijing 100193, China}
\affiliation{$^{3}$SKLSM, Institute of Semiconductors, Chinese Academy of Sciences, P.O.
Box 912, Beijing 100083, China}

\begin{abstract}
Despite many similarities between electronics and optics, the hopping of the
electron on a discrete atomic lattice gives rise to energy band
nonparabolicity and anisotropy. The crucial influences of this effect on
material properties and its incorporation into the continuum model have
received widespread attention in the past half century. Here we predict the
existence of a different effect due to the hopping of the electron across an
atomically sharp interface. For a general lattice, its influence on transport
could be equally important as the energy band nonparabolicity/anisotropy, but
cannot be incorporated into the continuum model. On the honeycomb lattice of
graphene, it leads to the breakdown of the conventional Klein tunneling -- one
of the exotic phenomena of relativistic particles -- and the onset of tilted
Klein tunneling. This works identifies a unique feature of the discrete atomic
lattice for transport, which is relevant for ballistic electronic devices at
high carrier densities.

\end{abstract}
\maketitle

The motion of electrons in solids shares many similarities to optics and
high-energy physics, but distinguishes itself by having a discrete atomic
lattice as the background. Exploring their similarities opens two research
fields: electron-optics in solids
\cite{SpectorAPL1990,SpectorAPL1990,DattaBook1995} (e.g., Veselago focusing
\cite{CheianovScience2007,LeeNatPhys2015,ChenScience2016,ZhangPRB2016,Zhang2DMater2017}%
, wave guiding \cite{WilliamsNatNano2011}, and quantum Goos-H\"{a}nchen effect
\cite{BeenakkerPRL2009,WuPRL2011}) and the condensed-matter analogs of
high-energy physics (e.g., Dirac fermions in graphene \cite{CastroRMP2009},
Majorana fermions in superconducting heterostructures
\cite{MourikScience2012,PergeScience2014}, and Weyl
\cite{LvNatPhys2015,XuNatPhys2015,XuScience2015,HirschbergerNatMater2016} and
Dirac semimetals \cite{LiuNatMater2014,LiuScience2014,XiongScience2015}).
Exploring the discrete nature of the lattice leads to the recent prediction of
unconventional fermions\ beyond high-energy physics
\cite{BradlynScience2016,BradlynNature2017} and the energy band
nonparabolicity and anisotropy. During the past half century, the crucial
influences of the energy band nonparabolicity/anisotropy on various material
properties have received widespread attention and intense efforts have been
devoted to incorporating them into continuum\textbf{ }$\mathbf{k}%
\cdot\mathbf{p}$ models -- the most versatile and physically transparent model
for studying material properties (see Refs.
\cite{PikusStrainBook1974,WinklerBook2003} for reviews).

In this work, we show that in addition to unconventional fermions and the
energy band nonparabolicity/anisotropy, the discrete lattice also gives rise
to a hitherto unidentified interfacial hopping effect. For a general lattice,
this effect could be equally important as the energy band nonparabolicity and
anisotropy, but cannot be reproduced by any continuum model, as opposed to the
energy band nonparabolicity/anisotropy. We begin with the concept of
interfacial hopping, first in a toy model -- a one-dimensional atomic chain --
and then in a general lattice model. Next we specialize to the honeycomb
lattice of graphene \cite{CastroRMP2009}, which provides an ideal platform for
ballistic electronic devices due to its exceptionally long ballistic length
\cite{WangScience2013,BaringhausNature2014}. The building block of these
devices is a junction (or interface), which exhibits optics-like behaviors
\cite{CheianovScience2007,LeeNatPhys2015,ChenScience2016,WilliamsNatNano2011,BeenakkerPRL2009,WuPRL2011,JiangPRL2013,ZhangPRB2016,Zhang2DMater2017}
and the celebrated Klein tunneling
\cite{KatsnelsonNatPhys2006,YoungNatPhys2009,StanderPRL2009,AllainEPJB2011,KatsnelsonBook2012}
-- the unimpeded penetration of relativistic particles through potential
barriers upon normal incidence. Using continuum models for graphene, the Klein
tunneling was derived by two seminal papers: one for smooth junctions
\cite{CheianovPRB2006} (i.e., junction width $\gg$ electron wavelength) and
the other for sharp junctions \cite{KatsnelsonNatPhys2006} (i.e., junction
width $\ll$ electron wavelength). Here we go one step further by showing that
when the junction becomes atomically sharp (i.e., junction width $\sim$ atomic
distance, as fabricated recently \cite{BaiPRB2018}), the interfacial hopping
leads to the breakdown of the conventional Klein tunneling and the onset of
\textit{tilted} Klein tunneling, i.e., perfect transmission at a tilted
incident angle $\theta_{\mathrm{c}}$ determined by the ration between the
atomic distance and the electron wavelength \footnote{The discrepancy between
the tight-binding model and the continuum model has been noted in earlier
works \cite{TangSSC2008,LiuPRB2012b}. However, they only consider normal
incidence and low doping (i.e., electron wavelength $\gg$ atomic distance), so
the discrepancy is very small (less than 0.5\%) and hence taken to be
negligible \cite{LiuPRB2012b}.}. This phenomenon cannot be reproduced by any
continuum model even if high-order terms are included into the continuum model
to reproduce exactly the nonparabolic and anisotropic energy band dispersion
and the spinor eigenstates over the whole Brillouin zone of graphene. Using
the experimentally demonstrated doping level \cite{DombrowskiPRL2017} and
junction potential \cite{BaiPRB2018}, the tilted incident angle $\theta
_{\mathrm{c}}\geq10%
\operatorname{{{}^\circ}}%
$ should be resolvable by multiprobe scanning tunneling microscopy
\cite{NakayamaAM2012,LiAFM2013,SettnesPRL2014} or transverse magnetic focusing
\cite{ChenScience2016}. We also discuss the feasibility of observing this
phenomenon in artificial \textquotedblleft photonic graphene\textquotedblright%
\ \cite{RechtsmanNP2012,PlotnikNM2013}. In addition to Klein tunneling, we
expect significant influence of the interfacial hopping effect on other exotic
transport phenomena in graphene and other high-mobility materials at high
carrier densities.

\begin{figure}[ptb]
\includegraphics[width=\columnwidth,clip]{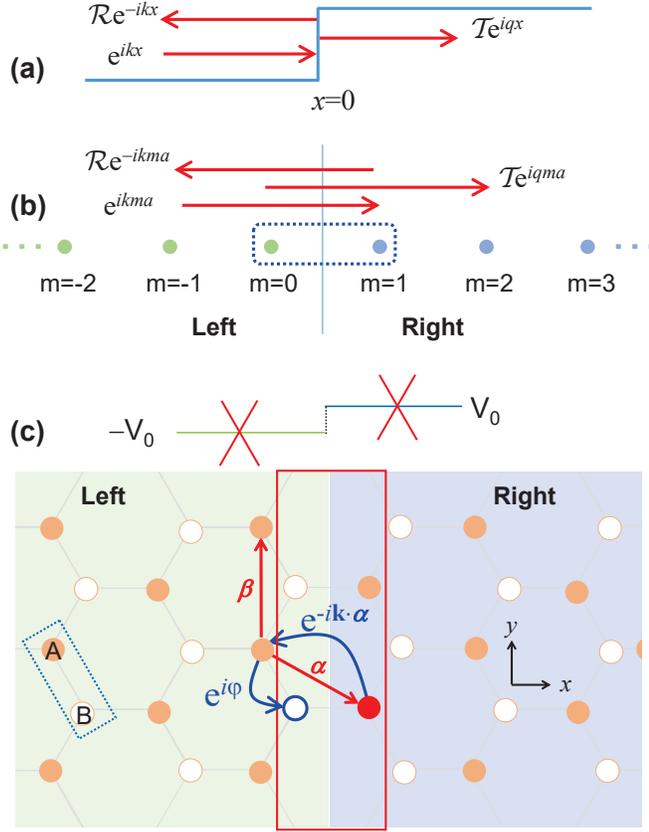}\caption{Transmission and
reflection across a sharp interface in (a) one-dimensional continuum model;
(b) one-dimensional lattice model with lattice spacing $a$; (c) honeycomb
lattice of graphene with two atoms ($A$ and $B$) in one unit cell, as marked
by the dashed rectangle. The solid rectangle marks the penetration region for
mode-matching.}%
\label{G_PNJ}%
\end{figure}

\textbf{Interfacial hopping effect}. We begin with the scattering of a
one-dimensional free electron with mass $m_{0}$ and incident energy $E_{F}$ by
a sharp interface to introduce the mode-matching method in the lattice model
\cite{AndoPRB1991,ZhangPRB2018} and the concept of interfacial hopping, with
$\hbar=1$ for brevity. In the continuum model, the interface is described by a
step-wise potential that vanishes on the left ($x<0$)\ and takes a constant
value $V$ on the right ($x>0$). As shown in Fig. \ref{G_PNJ}(a), a right-going
incident wave $e^{ikx}$ with $k\equiv\sqrt{2m_{0}E_{F}}$\ produces a
left-going reflection wave $\mathcal{R}e^{-ikx}$ and a right-going
transmission wave $\mathcal{T}e^{iqx}$, where $q\equiv\sqrt{2m_{0}(E_{F}-V)}$.
The reflection and transmission amplitudes $\mathcal{R}$ and $\mathcal{T}$ are
determined by the continuity of the scattering state and its derivative at the
interface:
\begin{subequations}
\label{CM_CE}%
\begin{align}
1+\mathcal{R} &  =\mathcal{T},\\
ik-ik\mathcal{R} &  =iq\mathcal{T}.
\end{align}
For a one-dimensional atomic chain with nearest-neighbor distance $a$ and
hopping energy $-t$ (with $t>0$), the interface is described by a constant
on-site energy $\varepsilon$ on the left ($m\leq0$)\ and $\varepsilon+V$ on
the right ($m\geq1$), as shown in Fig. \ref{G_PNJ}(b). The scattering state
emanating from an incident wave $e^{ikma}$ from the left satisfies the
Schr\"{o}dinger equation $(E_{F}-\varepsilon-V_{m})\Phi_{m}+t\Phi_{m+1}%
+t\Phi_{m-1}=0$, which reduces to
\end{subequations}
\begin{equation}
(E_{F}-\varepsilon)\Phi_{m}+t\Phi_{m+1}+t\Phi_{m-1}=0\ \ \ (m=-\infty
,\cdots,0)\label{SEQ_L}%
\end{equation}
on the left of the interface and%
\begin{equation}
(E_{F}-\varepsilon-V)\Phi_{m}+t\Phi_{m+1}+t\Phi_{m-1}=0\ (m=1,\cdots
,+\infty)\label{SEQ_R}%
\end{equation}
on the right. Equation\ (\ref{SEQ_L}) admits a right-going eigen-solution
$e^{ikma}$ (i.e., the incident wave) and a left-going eigen-solution
$e^{-ikma}$ (i.e., the reflection wave), where $k$ is determined by
$E_{F}=E(k)$ and $E(k)\equiv\varepsilon-2t\cos(ka)$ is the energy band
dispersion of the left region. For each $m$, Eq. (\ref{SEQ_L}) connects the
central site $\Phi_{m}$ to the neighboring sites $\Phi_{m-1}$ and $\Phi_{m+1}%
$, so the two eigen-solutions to Eq. (\ref{SEQ_L}) remain valid in the
penetration region [dashed box in Fig. \ref{G_PNJ}(b)]. In other words,
solving Eq. (\ref{SEQ_L}) gives the left-region scattering state $\Phi
_{m\leq1}=e^{ikma}+\mathcal{R}e^{-ikma}$, which penetrates into the site $m=1$
of the right region due to the hopping of the electron across the interface.
Similarly, Eq.\ (\ref{SEQ_R}) admits a right-going eigen-solution $e^{iqma}$
(i.e., the transmission wave) and a left-going eigen-solutions $e^{-iqma}$
(excluded by causality), where $q$ is determined by $E_{F}=V+E(q)$. Therefore,
solving Eq. (\ref{SEQ_R}) gives the right-region scattering state $\Phi
_{m\geq0}=\mathcal{T}e^{iqma}$, which penetrates into the site $m=0$ of the
left region due to the interfacial hopping. The reflection and transmission
amplitudes are determined by the continuity of $\Phi_{m}$ in the penetration
region ($m=0$ and $1$):
\begin{subequations}
\label{LM_CE}%
\begin{align}
1+\mathcal{R} &  =\mathcal{T},\label{LM_CE1}\\
e^{ika}+\mathcal{R}e^{-ika} &  =\mathcal{T}e^{iqa}.\label{LM_CE2}%
\end{align}

Compared with the continuum model, the lattice model reveals two distinct
effects: (i) Nonparabolicity/Anisotropy of the energy band $E(k)=\varepsilon
-2t\cos(ka)$, as opposed to the parabolic energy band $E(k)=k^{2}/(2m_{0})$ in
the continuum model; (ii) Interfacial hopping, which leads to the continuity
of the wave function at two \textit{different} sites $m=0$ and $1$ [Eq.
(\ref{LM_CE})], as opposed to the continuity of the wave function and its
derivative at the \textit{same} location $x=0$ [Eq. (\ref{CM_CE})] in the
continuum model. Both (i) and (ii) originate from the electron hopping on the
discrete lattice: the sum of the on-site energy $\varepsilon$ and the
nearest-neighbor hopping gives the energy band dispersion $E(k)=\varepsilon
+(-t)e^{ika}+(-t)e^{-ika}$, while the hopping from $m=0$ across the interface
onto $m=1$ gives rise to the continuity of the wave function at $m=0$ and $1$
and hence the propagation phases $e^{ika},e^{-ika},e^{iqa}$ of the incident,
reflection, and transmission waves in Eq. (\ref{LM_CE}). In the continuum
limit $ka\rightarrow0$, we can expand effect (i) up to the second order of
$ka$ and expand effect (ii) up to the first order to recover the continuum
model. For finite $ka$ (especially at high doping densities $ka\sim1$),
however, the high-order contributions from effects (i) and (ii) become equally important.

The results above can be extended to a general lattice with an arbitrary
number $M$ of orth-normalized atomic orbitals (associated with the same or
different atoms) in each unit cell. Let's label a unit cell by its center
location $\mathbf{r}$, use $|\mathbf{r},\tau\rangle$ for the $\tau$th orbital
in the $\mathbf{r}$-th unit cell, $\boldsymbol{\varepsilon}$ for the $M\times
M$ on-site energy, and $\mathbf{t}(\mathbf{r}-\mathbf{r}^{\prime})$ for the
$M\times M$ hopping matrix from the $\mathbf{r}$-th unit cell to the
$\mathbf{r}^{\prime}$-th unit cell. Under the Bloch basis $|\mathbf{k}%
,\tau\rangle=(1/\sqrt{N})\sum_{\mathbf{r}}e^{i\mathbf{k}\cdot\mathbf{r}%
}|\mathbf{r},\tau\rangle$ ($N$ is the total number of unit cells), the
Hamiltonian for a uniform lattice is $\mathbf{H}(\mathbf{k}%
)=\boldsymbol{\varepsilon}+\sum_{\mathbf{r}\neq\mathbf{0}}e^{i\mathbf{k}%
\cdot\mathbf{r}}\mathbf{t}(\mathbf{r})$. The eigenstate $e^{i\mathbf{k}%
\cdot\mathbf{r}}|u(\mathbf{k})\rangle$ is the product of the propagation phase
$e^{i\mathbf{k}\cdot\mathbf{r}}$ accompanying the electron hopping between
different unit cells and the $M$-component spinor $|u(\mathbf{k})\rangle$. In
a $d$-dimensional Bravais lattice, an interface is a $(d-1)$-dimensional
plane. For simplicity, we assume that this interface is defined by different
on-site energies (e.g., $\boldsymbol{\varepsilon}$ on one side of the
interface and $\boldsymbol{\varepsilon}+\mathbf{V}$ on the other side),
similar to the one-dimensional atomic chain. In this case, we can choose the
$d$ primitive vectors $\boldsymbol{\beta}_{1},\cdots,\boldsymbol{\beta}_{d}$
of the Bravais lattice such that $\boldsymbol{\beta}_{1},\cdots
,\boldsymbol{\beta}_{d-1}$ lie inside this interface. Given an incident wave
$e^{i\mathbf{k}_{\mathrm{I}}\cdot\mathbf{r}}|u_{\mathrm{I}}\rangle$, the
scattering state is $e^{i\mathbf{k}_{\mathrm{I}}\cdot\mathbf{r}}%
|u_{\mathrm{I}}\rangle+\sum_{\lambda}\mathcal{R}_{\lambda}e^{i\mathbf{k}%
_{\lambda,\mathrm{R}}\cdot\mathbf{r}}|u_{\lambda,\mathrm{R}}\rangle$ on one
side of the interface and $\sum_{\lambda}\mathcal{T}_{\lambda}e^{i\mathbf{k}%
_{\lambda,\mathrm{T}}\cdot\mathbf{r}}|u_{\lambda,\mathrm{T}}\rangle$ on the
other side, where $\lambda$ labels the different reflection and transmission
waves. If we consider electron hopping between neighboring unit cells only,
then
\end{subequations}
\begin{equation}
\mathbf{H}(\mathbf{k})=\boldsymbol{\varepsilon}+\sum_{s=1}^{d}e^{i\mathbf{k}%
\cdot\boldsymbol{\beta}_{s}}\mathbf{t}(\boldsymbol{\beta}_{s})\label{HK}%
\end{equation}
and the continuity equations across the interface gives
\begin{subequations}
\label{MM_GEN}%
\begin{align}
|u_{\mathrm{I}}\rangle+\sum_{\lambda}\mathcal{R}_{\lambda}|u_{\lambda
,\mathrm{R}}\rangle &  =\sum_{\lambda}\mathcal{T}_{\lambda}|u_{\lambda
,\mathrm{T}}\rangle,\label{MM_GEN1}\\
e^{i\mathbf{k}_{\mathrm{I}}\cdot\boldsymbol{\beta}_{d}}|u_{\mathrm{I}}%
\rangle+\sum_{\lambda}\mathcal{R}_{\lambda}e^{i\mathbf{k}_{\lambda,\mathrm{R}%
}\cdot\boldsymbol{\beta}_{d}}|u_{\lambda,\mathrm{R}}\rangle &  =\sum_{\lambda
}\mathcal{T}_{\lambda}e^{i\mathbf{k}_{\lambda,\mathrm{T}}\cdot
\boldsymbol{\beta}_{d}}|u_{\lambda,\mathrm{T}}\rangle.\label{MM_GEN2}%
\end{align}
Equation (\ref{MM_GEN}) is a direct generalization of Eq. (\ref{LM_CE}), e.g.,
Eqs. (\ref{MM_GEN1}) and (\ref{LM_CE1}) [Eqs. (\ref{MM_GEN2}) and
(\ref{LM_CE2})] come from the continuity equation on the unit cell at the left
(right) neighborhood of the interface. Therefore, the electron hopping on the
discrete lattice leads to two equally important effects: (i) The hopping
between neighboring unit cells gives rise to the phase factors $e^{i\mathbf{k}%
\cdot\boldsymbol{\beta}_{s}}$ in the lattice Hamiltonian Eq. (\ref{HK}) and
hence energy band nonparabolicity/anisotropy. (ii) The hopping across the
interface leads to wave function continuity at two \textit{different} unit
cells across the interface and hence the propagation phase factors
$e^{i\mathbf{k}_{\mathrm{I}}\cdot\boldsymbol{\beta}_{d}}$, $e^{i\mathbf{k}%
_{\lambda,\mathrm{R}}\cdot\boldsymbol{\beta}_{d}}$, $e^{i\mathbf{k}%
_{\lambda,\mathrm{T}}\cdot\boldsymbol{\beta}_{d}}$ in Eq. (\ref{MM_GEN}) --
the interfacial hopping effect. For many years, the influence of (i) on
various material properties have received widespread attention and intense
efforts have been devoted to incorporating them into continuum\textbf{
}$\mathbf{k}\cdot\mathbf{p}$ models \cite{PikusStrainBook1974,WinklerBook2003}%
. By contrast, effect (ii) is unique to the lattice model and we are not aware
of any quantitative discussion about its influence on mesoscopic transport.

\textbf{Tilted Klein tunneling across atomically sharp graphene junction.}
Atomically-sharp graphene junction was fabricated \cite{BaiPRB2018} and
studied numerically \cite{LiuPRB2012b,LogemannPRB2015}. Here we use this
system to demonstrate the interfacial hopping effect in transport. As shown in
Fig. \ref{G_PNJ}(c), graphene has a triangular Bravais lattice, with two
carbon atoms (marked by $A$ and $B$) in one unit cell and the carbon-carbon
bond length $a=0.142$ nm. Uniform graphene described by the tight-binding
model with nearest-neighbor hopping admits a conduction band with dispersion
$t|f(\mathbf{k})|$ and a valence band with dispersion $-t|f(\mathbf{k})|$,
where $t=2.7$ eV is the nearest-neighbor hopping energy \cite{CastroRMP2009},
$f(\mathbf{k})=-(1+e^{-i\mathbf{k}\cdot\boldsymbol{\alpha}}+e^{i\mathbf{k}%
\cdot\boldsymbol{\beta}})$, and $\boldsymbol{\alpha}$,\textbf{ }%
$\boldsymbol{\beta}$ are primitive Bravais vectors, as indicated by the red
arrows in Fig. \ref{G_PNJ}(c). Since $f(\mathbf{k})$ vanishes at two
inequivalent points $\mathbf{K}=(2\pi/3a)(1,1/\sqrt{3})$ and $\mathbf{K}%
^{\prime}=(2\pi/3a)(1,-1/\sqrt{3})$ in the reciprocal space, the energy bands
of graphene are approximately described by a Dirac-like continuum-model near
$\mathbf{K}$ and $\mathbf{K}^{\prime}$. An electron with momentum $\mathbf{k}$
and energy $E_{F}$ has a wave function $e^{i\mathbf{k}\cdot\mathbf{r}%
}[1,e^{i\varphi}]^{T}$ (spinors always normalized to $\sqrt{2}$ as a
convention), where $e^{i\varphi}\equiv tf^{\ast}(\mathbf{k})/E_{F}$ is a phase
factor. To eliminate unwanted intervalley scattering \cite{UrbanPRB2011} and
focus on the interfacial hopping, we take the junction along the zigzag
direction (defined as the $y$ axis for clarity), then an incident wave in one
valley only produces reflection and transmission waves in the same valley. The
junction potential is described by a constant on-site energy $-V_{0}$
($+V_{0}$) for all the carbon sites in the left (right) region
\cite{ZhangPRB2017,ZhangPRB2018} [see the upper panel of Fig. \ref{G_PNJ}(c)].
The Fermi energy $E_{F}$ determines the dimensionless doping level
$\varepsilon_{\mathrm{L}}\equiv(E_{F}+V_{0})/t$ for the left region and
$\varepsilon_{\mathrm{R}}\equiv(E_{F}-V_{0})/t$ for the right region: a
positive (negative) doping level corresponds to electron or N (hole or P)
doping, e.g., positive (negative) $\varepsilon_{\mathrm{L}}$ and
$\varepsilon_{\mathrm{R}}$ describe N-N (P-P) junction, while positive
$\varepsilon_{\mathrm{L}}$ and negative $\varepsilon_{\mathrm{R}}$ or vice
versa describe P-N junctions.

We consider the scattering of an incident wave with energy $E_{F}$ and
momentum $\mathbf{k}_{\mathrm{I}}=(k_{x},k_{y})$ on the left region. Given
$E_{F}$ and $k_{y}$, $k_{x}$ is uniquely determined by $|\varepsilon
_{\mathrm{L}}|=|f(\mathbf{k}_{\mathrm{I}})|$ and the requirement that the
group velocity of the incident wave has a positive projection along the $+x$
axis. The translational invariance along the interface ($y$ axis) leads to the
conservation of $k_{y}$, so the incident wave produces a left-going reflection
wave with momentum $\mathbf{k}_{\mathrm{R}}=(-k_{x},k_{y})$ [since
$|f(\mathbf{k})|$ is an even function of $k_{x}$] and a right-going
transmission wave with momentum $\mathbf{k}_{\mathrm{T}}=(k_{x,\mathrm{T}%
},k_{y})$, where $k_{x,\mathrm{T}}$ is uniquely determined by $\left\vert
\varepsilon_{\mathrm{R}}\right\vert =\left\vert f(\mathbf{k}_{\mathrm{T}%
})\right\vert $ and the requirement that the group velocity of the
transmission wave has a positive projection along the $+x$ axis. In addition
to these traveling waves, graphene also supports ideal evanescent waves.
Fortunately, the latter do not affect the scattering across a sharp zigzag
junction and hence can be discarded \cite{ZhangPRB2018} as long as we employ
the wave function continuity in the penetration region [red solid square in
Fig. \ref{G_PNJ}(c)]. On the left, including the penetration region, the
scattering state
\end{subequations}
\[
|\Phi(\mathbf{r})\rangle=e^{i\mathbf{k}_{\mathrm{I}}\cdot\mathbf{r}}%
\begin{bmatrix}
1\\
e^{i\varphi_{\mathrm{I}}}%
\end{bmatrix}
+\mathcal{R}e^{i\mathbf{k}_{\mathrm{R}}\cdot\mathbf{r}}%
\begin{bmatrix}
1\\
e^{i\varphi_{\mathrm{R}}}%
\end{bmatrix}
\]
is the sum of the incident wave and the reflection wave. On the right,
including the penetration region, the scattering state
\[
|\Phi(\mathbf{r})\rangle=\mathcal{T}e^{i\mathbf{k}_{\mathrm{T}}\cdot
\mathbf{r}}%
\begin{bmatrix}
1\\
e^{i\varphi_{\mathrm{T}}}%
\end{bmatrix}
\]
is the transmission wave, where $e^{i\varphi_{\mathrm{I}}}\equiv f^{\ast
}(\mathbf{k}_{\mathrm{I}})/\varepsilon_{\mathrm{L}}$, $e^{i\varphi
_{\mathrm{R}}}\equiv f^{\ast}(\mathbf{k}_{\mathrm{R}})/\varepsilon
_{\mathrm{L}}$, and $e^{i\varphi_{\mathrm{T}}}\equiv f^{\ast}(\mathbf{k}%
_{\mathrm{T}})/\varepsilon_{\mathrm{R}}$. With the origin $\mathbf{r}=0$ of
the coordinate set on the $A$ atom [filled red circle in Fig. \ref{G_PNJ}(c)]
in the penetration region, the continuity of the scattering state on this atom
gives $1+\mathcal{R}=\mathcal{T}$, while the continuity of the scattering
state on the $B$ atom [empty blue circle in Fig. \ref{G_PNJ}(c)] gives
$e^{i(\varphi_{\mathrm{I}}-\mathbf{k}_{\mathrm{I}}\cdot\boldsymbol{\alpha}%
)}+\mathcal{R}e^{i(\varphi_{\mathrm{R}}-\mathbf{k}_{\mathrm{R}}\cdot
\boldsymbol{\alpha})}=\mathcal{T}e^{i(\varphi_{\mathrm{T}}-\mathbf{k}%
_{\mathrm{T}}\cdot\boldsymbol{\alpha})},$ where $\boldsymbol{\alpha}$ is the
primitive vector of the graphene lattice [see the red arrow Fig.
\ref{G_PNJ}(c)]. Here $e^{i\varphi_{\lambda}}$ ($\lambda=\mathrm{I,R,T}$)
accounts for the relative phase between $B,A$ atoms in the same unit cell,
while $e^{-i\mathbf{k}_{\lambda}\cdot\boldsymbol{\alpha}}$ are propagation
phase factors accompanying the electron hopping across the interface [see the
blue arrow in Fig. \ref{G_PNJ}(c)], i.e., interfacial hopping, which is absent
in \textit{any} continuum model. The continuity of the wave function in the
penetration region can be written concisely as%
\begin{equation}%
\begin{bmatrix}
1\\
e^{i(\varphi_{\mathrm{I}}-\mathbf{k}_{\mathrm{I}}\cdot\boldsymbol{\alpha})}%
\end{bmatrix}
+\mathcal{R}%
\begin{bmatrix}
1\\
e^{i(\varphi_{\mathrm{R}}-\mathbf{k}_{\mathrm{R}}\cdot\boldsymbol{\alpha})}%
\end{bmatrix}
=\mathcal{T}%
\begin{bmatrix}
1\\
e^{i(\varphi_{\mathrm{T}}-\mathbf{k}_{\mathrm{T}}\cdot\boldsymbol{\alpha})}%
\end{bmatrix}
,\label{MM_CD}%
\end{equation}
which is a special case of Eq. (\ref{MM_GEN}). The condition for zero
reflection or equivalently perfect transmission (i.e., $\mathcal{R}=0$ and
$\mathcal{T}=1$) is%
\begin{equation}
\frac{e^{i\varphi_{\mathrm{I}}}}{e^{i\varphi_{\mathrm{T}}}}=\frac
{e^{i\mathbf{k}_{\mathrm{I}}\cdot\boldsymbol{\alpha}}}{e^{i\mathbf{k}%
_{\mathrm{T}}\cdot\boldsymbol{\alpha}}}.\label{MC2}%
\end{equation}
The left hand side is affected by the energy band nonparabolicity and
anisotropy, while the right hand side accounts for the interfacial hopping.
Due to the mirror symmetry of the junction about the $x$ axis [see Fig.
\ref{G_PNJ}(c)], the reflection probability $|\mathcal{R}|^{2}$ is an even
function of $k_{y}$, so we need only consider the $\mathbf{K}$ valley (i.e.,
$k_{y}>0$) and obtain
\begin{equation}
\cos\frac{\sqrt{3}k_{y}a}{2}=\frac{\sqrt{1-\varepsilon_{\mathrm{L}}%
\varepsilon_{\mathrm{R}}}}{2}.\label{KY}%
\end{equation}
When the doping level is well below the Van Hove singularity ($|\varepsilon
_{\mathrm{L},\mathrm{R}}|\ll1$), we can use $|f(\mathbf{k})|\approx\alpha
_{x}t|\mathbf{q}|$ and $|q_{y}|a\ll1$ with $\mathbf{q}\equiv\mathbf{k}%
-\mathbf{K}$ being the reduced momentum to obtain $q_{y}\alpha_{x}%
\approx\varepsilon_{\mathrm{L}}\varepsilon_{\mathrm{R}}/2$, so perfect
transmission occurs at a tilted incident angle \footnote{The incident
(transmission) angle is the polar angle of the incident (transmission) group
velocity about the $x$ axis. In general, this differs from that of the
incident (transmission) momentum. When the doping level lies well below the
Van Hove singularity of the graphene energy band, however, we can neglect this
small difference, so that the polar angle of the reduced momentum is parallel
(anti-parallel) to the polar angle of the group velocity.}
\begin{equation}
\theta_{\mathrm{c}}\approx\frac{\varepsilon_{\mathrm{R}}}{2}\approx
\mathrm{sgn}(\varepsilon_{\mathrm{R}})\frac{3\pi a}{2\lambda_{\mathrm{R}}%
},\label{THETA_I}%
\end{equation}
where $\lambda_{\mathrm{R}}$\ is the Fermi wavelength in the right (i.e.,
transmission) region. The corresponding transmission angle $\theta
_{\mathrm{T}}\approx\varepsilon_{\mathrm{L}}/2$ is obtained from the Snell's
law $\varepsilon_{\mathrm{R}}\sin\theta_{\mathrm{T}}=\varepsilon_{\mathrm{L}%
}\sin\theta_{\mathrm{c}}$, which follows from the conservation of $q_{y}$
across the junction.

\begin{figure}[ptb]
\includegraphics[width=\columnwidth,clip]{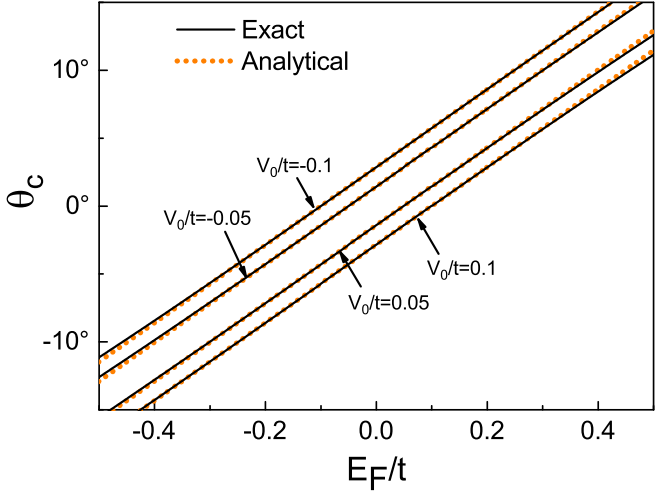}\caption{Critical
incident angle $\theta_{\mathrm{c}}$ for zero reflection (or perfect
transmission) as a function of the Fermi energy for different junction
potentails: solid lines for exact results and dotted lines for the analytical
expression Eq. (\ref{THETA_I}). The region $|E_{F}|<V_{0}$ corresponds to P-N
junctions, $E_{F}>|V_{0}|$ corresponds to N-N junctions, and $E_{F}<-|V_{0}|$
corresponds to P-P junctions.}%
\label{G_ANGLE}%
\end{figure}

Equation (\ref{THETA_I}) is valid for P-N, N-N, and P-P junctions, which
correspond to $|E_{F}|<V_{0}$, $E_{F}>|V_{0}|$, and $E_{F}<-|V_{0}|$,
respectively. As shown in Fig. \ref{G_ANGLE}, the analytical expression Eq.
(\ref{THETA_I}) agrees well with the exact numerical results over a wide
doping level well beyond the linear Dirac regime, as long as the doping level
lies below the Van Hove singularity. In contrast to the well-known Klein
tunneling at normal incidence ($\theta_{\mathrm{c}}=0$)
\cite{KatsnelsonNatPhys2006,AllainEPJB2011}, Eq. (\ref{THETA_I}) shows that
zero reflection or equivalently perfect transmission across an atomically
sharp graphene junction requires a \textit{tilted} incidence angle
$\theta_{\mathrm{c}}$ that is uniquely determined by the dimensionless doping
level $\varepsilon_{\mathrm{R}}$ on the transmission region. This phenomenon
becomes very important when the Fermi wave length $\lambda_{\mathrm{R}}$ in
the transmission region is comparable to the atomic distance $a$. In Figs.
\ref{G_ANGLE}(a) and \ref{G_ANGLE}(b), we show the contour plot for the
reflection probability $|\mathcal{R}|^{2}$ as a function of the Fermi energy
$E_{F}$ and the incidence angle for two different junction potentials
$V_{0}=0.05t$ and $V_{0}=0.1t$. The tilted incident angle $\theta_{\mathrm{c}%
}$ (as marked by the white solid lines) shows pronounced deviation from normal
incidence. Next we show that the interfacial hopping plays an important role
in this phenomenon.

\begin{figure}[ptb]
\includegraphics[width=\columnwidth,clip]{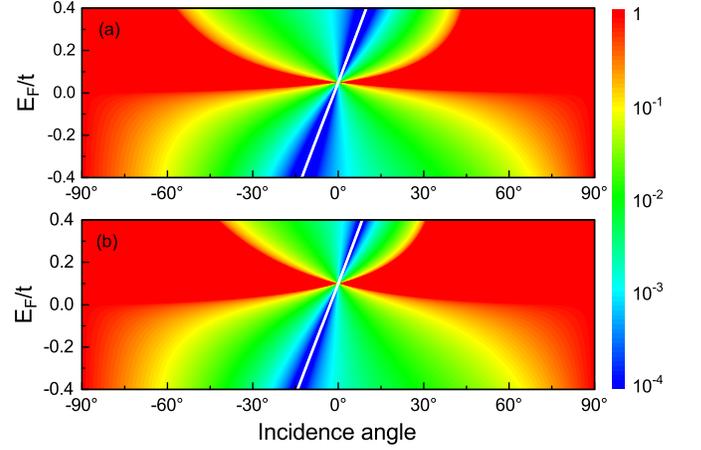}\caption{Reflection
probability as functions of the Fermi energy and the incident angle for
different junction potentials: (a) $V_{0}=0.05t$ and (b) $V_{0}=0.1t$. The
white solid line mark the critical incident angle $\theta_{\mathrm{c}}$ for
zero reflection or equivalently perfection transmission.}%
\label{G_CONTOUR}%
\end{figure}

\textbf{Breaking pseudospin conservation by interfacial hopping.} To
understand the physical origin of the \textit{tilted} Klein tunneling in the
lattice model, we recall that in the linear continuum model
\cite{KatsnelsonNatPhys2006,AllainEPJB2011}, the Hamiltonian of uniform
graphene is $H_{0}=v_{F}\boldsymbol{\sigma}\cdot\mathbf{p}$ and the
eigenstates are $e^{i\mathbf{q}\cdot\mathbf{r}}[1,e^{i\phi}]^{T}$, where
$\phi$ is the azimuth angle of the reduced momentum $\mathbf{q=k}-\mathbf{K}$.
The spinor $[1,e^{i\phi}]^{T}$ describes a pseudospin lying in the $xy$ plane
with an azimuth angle $\phi$. In this model, the scattering state across the
junction is given by $e^{i\mathbf{q}_{\mathrm{I}}\cdot\mathbf{r}}%
[1,e^{i\phi_{\mathrm{I}}}]+\mathcal{R}e^{i\mathbf{q}_{\mathrm{R}}%
\cdot\mathbf{r}}[1,e^{i\phi_{\mathrm{R}}}]$ on the left side and
$\mathcal{T}e^{i\mathbf{q}_{\mathrm{T}}\cdot\mathbf{r}}[1,e^{i\phi
_{\mathrm{T}}}]$ on the right side, where $\mathbf{q}_{\lambda}\equiv
\mathbf{k}_{\lambda}-\mathbf{K}$ ($\lambda=\mathrm{I,R,T}$). The continuity of
the scattering state across the interface at $x=0$ gives%
\[%
\begin{bmatrix}
1\\
e^{i\phi_{\mathrm{I}}}%
\end{bmatrix}
+\mathcal{R}%
\begin{bmatrix}
1\\
e^{i\phi_{\mathrm{R}}}%
\end{bmatrix}
=\mathcal{T}%
\begin{bmatrix}
1\\
e^{i\phi_{\mathrm{T}}}%
\end{bmatrix}
,
\]
which resembles its lattice counterpart [Eq. (\ref{MM_CD})] ($e^{i\phi
_{\lambda}}$ approaches $e^{i\varphi_{\lambda}}$ when $\mathbf{k}_{\lambda}$
approaches $\mathbf{K}$), but does not carry the interfacial hopping phase
$e^{-i\mathbf{k}_{\lambda}\cdot\boldsymbol{\alpha}}$. As a result, in the
continuum model, perfect transmission requires $e^{i\phi_{\mathrm{I}}%
}=e^{i\phi_{\mathrm{T}}}$ or equivalently the pseudospin of the incident wave
being parallel to that of the transmission wave, which is satisfied upon
normal incidence (i.e., $\phi_{\mathrm{I}}=\phi_{\mathrm{T}}=0$). By contrast,
in the lattice model, the perfect transmission condition requires Eq.
(\ref{MC2}), which is not satisfied upon normal incidence (e.g.,
$\varphi_{\mathrm{I}}=\varphi_{\mathrm{R}}=0$). This is because in the
continuum model, the continuity equation for the scattering wave function
occurs at the \textit{same} location -- the interface, while in the lattice
model, the continuity equation occurs at two \textit{different} unit cells
across the interface -- the interfacial hopping effect. Despite the same
momentum conservation along the interface in both the continuum model and the
lattice model, the unique interfacial hopping effect in the lattice model
leads to the breakdown of the pseudospin conservation and hence the
conventional Klein tunneling.

In Eq. (\ref{MC2}), the nonparabolicity and anisotropy of the graphene energy
band influences the values of the momenta $\mathbf{k}_{\mathrm{I}}$,
$\mathbf{k}_{\mathrm{R}},\mathbf{k}_{\mathrm{T}}$, and the spinor phases
$e^{i\varphi_{\mathrm{I}}},e^{i\varphi_{\mathrm{R}}},e^{i\varphi_{\mathrm{T}}%
}$. This effect can be well incorporated by using more sophisticated continuum
models, e.g., by adding higher-order terms or using more energy bands in the
$\mathbf{k}\cdot\mathbf{p}$ Hamiltonian. By contrast, the interfacial hopping
phases $e^{i\mathbf{k}_{\mathrm{I}}\cdot\boldsymbol{\alpha}}$ and
$e^{i\mathbf{k}_{\mathrm{T}}\cdot\boldsymbol{\alpha}}$ in the continuity
equation cannot be incorporated into the continuum model. If we fully
incorporate the former by using the exact momenta $\mathbf{k}_{\mathrm{I}}$,
$\mathbf{k}_{\mathrm{R}},\mathbf{k}_{\mathrm{T}}$, and spinor phases
$e^{i\varphi_{\mathrm{I}}},e^{i\varphi_{\mathrm{R}}},e^{i\varphi_{\mathrm{T}}%
}$ from the tight-binding model, but neglect the latter by replacing the right
hand side of Eq. (\ref{MC2}) with unity, then perfect transmission occurs at
\[
\cos\frac{\sqrt{3}k_{y}a}{2}=\frac{\sqrt{1+\varepsilon_{\mathrm{L}}%
\varepsilon_{\mathrm{R}}}}{2},
\]
as opposed to Eq. (\ref{KY}). When the doping level is well below the Van Hove
singularity, the critical incident angle $\theta_{\mathrm{c}}$ would be
$\theta_{\mathrm{c}}\approx-\varepsilon_{\mathrm{R}}/2$, which is
\textit{opposite} to the correct result in Eq. (\ref{THETA_I}). This suggests
that the interfacial hopping dominates over the nonparabolicity/anisotropy of
the energy band. As the interfacial hopping is unique to the lattice model,
this phenomenon differs qualitatively from existing electron-optics phenomena
\cite{CheianovScience2007,LeeNatPhys2015,ChenScience2016,WilliamsNatNano2011,BeenakkerPRL2009,WuPRL2011,ZhaiPRB2008,LiNL2017,NguyenPRB2018,ZhangPRB2018b}
describable by continuum models.

\textbf{Discussions. }The critical incidence angle $\theta_{\mathrm{c}}%
\approx\varepsilon_{\mathrm{R}}/2$ is completely determined by the
dimensionless doping level $\varepsilon_{\mathrm{R}}=(E_{F}-V_{0})/t$ in the
transmission region, while the corresponding transmission angle $\theta
_{\mathrm{T}}\approx\varepsilon_{\mathrm{L}}/2$ is completely determined by
$\varepsilon_{\mathrm{L}}=(E_{F}+V_{0})/t$ in the incidence region. With the
recently fabricated atomically sharp graphene P-N junction as an experimental
platform \cite{BaiPRB2018}, both $\theta_{\mathrm{c}}$ and $\theta
_{\mathrm{T}}$ can be tuned via the standard electrostatic doping technique
\cite{AhnRMP2006}, which has been widely used for graphene and other
two-dimensional materials. Experimentally, the potential across an atomically
sharp junction can reach $V_{0}\approx0.1t$ (see Ref. \cite{BaiPRB2018}) and
the uniform doping level can reach the Van Hove singularity (i.e., $|E_{F}%
|=t$) (see Ref. \cite{DombrowskiPRL2017}). For a rough estimate, we take
$V_{0}=-0.1t$ and $E_{F}=0.3t$, then $\varepsilon_{\mathrm{R}}=0.4$ and the
critical incident angle is $\theta_{\mathrm{c}}\approx11%
\operatorname{{{}^\circ}}%
$. This strong deviation of $\theta_{\mathrm{c}}$ from normal incidence (see
Fig. \ref{G_CONTOUR}) should be resolvable in multiprobe scanning tunneling
microscopy \cite{SettnesPRL2014} (see Refs. \cite{NakayamaAM2012,LiAFM2013}
for recent reviews) or in transverse magnetic focusing measurements
\cite{ChenScience2016}.

Another candidate experimental platform to observe this phenomenon is the
\textquotedblleft photonic graphene\textquotedblright%
\ \cite{RechtsmanNP2012,PlotnikNM2013}, an array of evanescently coupled
waveguides arranged in a honeycomb-lattice configuration
\cite{Bahat-TreidelPRL2010,PlotnikNM2013} (see Ref. \cite{PoliniNatNano2013}
for a review). Each waveguide has a single bound state, so the diffraction of
light in this structure is described by the same tight-binding model as
graphene. Here\ each waveguide mode corresponds to an atom and the tunneling
between neighboring waveguides corresponds to the hopping between neighboring
atoms. Importantly, the structure of the photonic lattice can be designed at
will and is not subject to structural defects or absorbate contamination, so
photonic graphene provides a platform for graphene physics not easily
accessible otherwise. In particular, the distance between neighboring
waveguides (corresponding to the atomic distance $a$ in the honeycomb lattice)
can be easily made comparable to the wavelength $\lambda_{\mathrm{R}}$ of the
photons, so that the critical incident angle $\theta_{\mathrm{c}}\sim
a/\lambda_{\mathrm{R}}$ deviates strongly from normal incidence, making the
observation of this phenomenon feasible.

To summarize, for many years, the crucial roles of the energy band
nonparabolicity/anisotropy and their incorporation into the continuum model
has attracted a lot of attention. Here our work have identified the
interfacial hopping effect as a missing ingredient that could be equally
important as the energy band nonparabolicity/anisotropy for transport across atomically sharp interfaces, but it cannot be
incorporated into the continuum model. The interfacial hopping shares the same physical origin as the the energy band
nonparabolicity/anisotropy, so it is universal for any solid-state materials. Specializing to the honeycomb
lattice of graphene reveals the breakdown of the well-known Klein tunneling
and the onset of tilted Klein tunneling. We expect significant influence of
the interfacial hopping effect on mesoscopic transport in other high-mobility
materials at high carrier densities.

\textbf{Acknowledgements. }This work was supported by the National Key R$\&$D
Program of China (Grant No. 2017YFA0303400), the NSFC (Grants No. 11504018, 11774021, and 11434010), and the NSFC program for \textquotedblleft Scientific
Research Center\textquotedblright\ (Grant No. U1530401). We acknowledge the computational support from the Beijing
Computational Science Research Center (CSRC).

\textbf{Author contributions}. S.H.Z., W.Y., and K.C. conceived the idea.
S.H.Z. and W.Y. formulated the theory, carried out the calculations, and wrote
the paper. All authors commented on the manuscript.

\textbf{Competing interests:} The authors declare no competing interests.

\end{document}